\documentclass[a4paper,11pt]{article}
\pdfoutput=1 

\usepackage{jheppub} 
\usepackage{amsmath,amssymb,amsthm,amscd,graphicx}
\usepackage{slashed}
\usepackage[notref,notcite]{}
\input epsf.sty

\theoremstyle{definition}

\newcommand{\CF}{{\cal F}}

\newcommand{\CI}{{\cal I}}

\newcommand{\CL}{{\cal L}}

\newcommand{\CO}{{\cal O}}

\newcommand{\CR}{{\cal R}}

\def\IR{{\mathbb R}}

\newcommand{\re}{{\rm e}}
\newcommand{\ri}{{\rm i}}
\newcommand{\rd}{{\rm d}}

\newcommand{\be}{\begin{equation}}
\newcommand{\ee}{\end{equation}}
\newcommand{\ba}{\begin{aligned}}
\newcommand{\ea}{\end{aligned}}
\newcommand{\ben}{\begin{eqnarray}\displaystyle}
\newcommand{\een}{\end{eqnarray}}

\newcommand{\sectiono}[1]{\section{#1}\setcounter{equation}{0}}

\newdimen\tableauside\tableauside=1.0ex
\newdimen\tableaurule\tableaurule=0.4pt
\newdimen\tableaustep
\def\phantomhrule#1{\hbox{\vbox to0pt{\hrule height\tableaurule width#1\vss}}}
\def\phantomvrule#1{\vbox{\hbox to0pt{\vrule width\tableaurule height#1\hss}}}
\def\sqr{\vbox{%
  \phantomhrule\tableaustep
  \hbox{\phantomvrule\tableaustep\kern\tableaustep\phantomvrule\tableaustep}%
  \hbox{\vbox{\phantomhrule\tableauside}\kern-\tableaurule}}}
\def\squares#1{\hbox{\count0=#1\noindent\loop\sqr
  \advance\count0 by-1 \ifnum\count0>0\repeat}}
\def\tableau#1{\vcenter{\offinterlineskip
  \tableaustep=\tableauside\advance\tableaustep by-\tableaurule
  \kern\normallineskip\hbox
    {\kern\normallineskip\vbox
      {\gettableau#1 0 }%
     \kern\normallineskip\kern\tableaurule}%
  \kern\normallineskip\kern\tableaurule}}
\def\gettableau#1{\ifnum#1=0\let\next=\null\else
\squares{#1}\let\next=\gettableau\fi\next}

\newcommand{\figref}[1]{Fig.~\protect\ref{#1}}

\title{\Huge{\boldmath Three roads to the energy gap}}

\author{Marcos Mari\~no and Tom\'as Reis}

\affiliation{D\'epartement de Physique Th\'eorique et Section de Math\'ematiques\\
Universit\'e de Gen\`eve, Gen\`eve, CH-1211 Switzerland}

\emailAdd{Marcos.Marino@unige.ch, Tomas.Reis@unige.ch} 

\abstract{We determine analytically the energy gap at weak coupling in the attractive multi-component Gaudin--Yang model, 
an integrable model which describes interacting fermions in one dimension with $\kappa$ components. 
We use three different methods. The first one is based on a direct analysis of the Bethe ansatz equations. The second method uses 
the theory of resurgence and the large order behavior of the perturbative series for the ground state energy. 
The third method is based on a renormalization group analysis. The three methods lead to the same answer, providing in this way 
a non-trivial test of the ideas of resurgence and renormalons as applied to non-relativistic many-body systems.}    

\begin{document}
\maketitle
\flushbottom
 
\sectiono{Introduction}

One of the most important non-perturbative effects in quantum theory is the energy gap of 
many-fermion systems with an attractive interaction. This gap, which is exponentially small in the coupling constant, 
is a universal feature of these systems, and it is at the origin of conventional superconductivity. However, being a 
non-perturbative effect, it is not easy to compute. One possibility is to use BCS-like mean field theory, which 
provides an approximate expression for the gap. Another possibility is to use renormalization group methods. In 
exactly solvable models, one can often calculate the gap exactly, and this provides a useful test of approximate 
methods. 

Recently, it has been pointed out that the leading behavior of the energy gap at weak coupling can 
be obtained from the large order behavior 
of the perturbative expansion for the ground state energy \cite{mr1,mr2,mr3}. 
This is an example of the general connection between perturbative series 
and non-perturbative effects first pointed out in quantum mechanics in \cite{lam,bw2}. The relationship 
between perturbative and non-perturbative 
sectors has evolved into a general framework to understand 
non-perturbative effects in mathematics and physics, sometimes called the theory of resurgence (see \cite{ss, mmlargen, abs,du-review} for reviews). 
In the case of many-fermion systems with an attractive interaction, 
it has been argued in \cite{mr1, mr2} that the energy gap is structurally very similar to a 
renormalon effect \cite{beneke} 
in an asymptotically free theory. Therefore, one can use renormalon techniques, like all-order calculations based on particular families of diagrams, 
to obtain information on the energy gap. These ideas were tested in two integrable models: the Gaudin--Yang model \cite{mr1,mr2} and 
the one-dimensional Hubbard model \cite{mr3}, in the case of attractive fermions with two components. 

In this paper we consider the Gaudin--Yang model with $\kappa$ components and $SU(\kappa)$ symmetry, which was briefly adressed in \cite{mr1}. This model is integrable \cite{sutherland,taka} and has many interesting features. First of all, it might be relevant to the study of ultracold 
atoms with higher hyperfine spin in one-dimensional traps (see e.g. \cite{guan-review,lecheminant2} and references therein). In addition, it displays new qualitative phenomena: the ground state consists of bound states of 
$\kappa$ elementary fermions, which generalize the familiar Cooper pairs occurring when $\kappa=2$ (when $\kappa=3$, these 
bound states are sometimes called ``trions"). More generally, one finds bound states of $1\le n\le \kappa$ fermions, leading to a rich phase structure. From a more theoretical point of view, this model might be an interesting testing ground for approximations based on a large number of components (``large $N$"). 

Here we are interested on the non-perturbative aspects of the model, 
and for this reason we will focus on its energy gap, in the weak coupling regime. In principle, 
the energy gap can be determined from the 
Bethe ansatz solution, as first pointed out in \cite{ko} in the case $\kappa=2$. However, an analytic calculation at weak coupling has 
not been performed for $\kappa>2$, since it requires a detailed study of the Bethe ansatz equations similar to 
what was done in \cite{mr1}. Our first result is then a formula for the energy gap, at next-to-leading order in the coupling constant, including the precise, $\kappa$-dependent prefactor. 

According to the conjecture of \cite{mr1}, we expect the energy gap to control 
the large order behavior of the perturbative series for the ground state energy. Such a connection was established numerically in 
\cite{mr1} for $\kappa=2$, and  
we check in detail that this connection persists for general $\kappa$. This provides a precision 
test of the ideas of resurgence, since the growth of perturbation theory at next-to-leading order in the number of loops ``predicts" the dependence of the gap on the coupling at next-to-leading order. In fact, in the case of general $\kappa$, we first found this dependence by looking at the large order behavior of the perturbative series, and only later we verified it with the Bethe ansatz calculation presented here. 

As discussed in \cite{mr1,mr3}, the non-perturbative scale leading to the energy gap can be regarded as a renormalon effect. 
It is due to special classes of diagrams which diverge 
factorially after integration over the momenta. As shown in \cite{mr1}, the ring diagrams dominating at large $\kappa$ are renormalon diagrams, and they lead to 
the right value for the leading order dependence of the gap on the coupling constant. However, we check explicitly that they fail to capture the next-to-leading 
dependence, which is to be expected since this dependence is subleading in the $1/\kappa$ expansion. 
It is well-known however that, in asymptotically free quantum field theories, the coupling constant dependence of the non-perturbative scale 
can be determined by a renormalization group (RG) analysis. The leading, exponential dependence of the non-perturbative scale is a 
one-loop effect, while the next-to-leading dependence requires knowledge of the beta function at two loops (see e.g. \cite{beneke}). 
In many-fermion systems, a similar argument linking the 
energy gap to an RG analysis was presented by Larkin and Sak in \cite{ls}, again in the case $\kappa=2$. In view of this connection, the results that we 
have obtained for the gap predict the form of the two-loop beta function of the Gaudin--Yang model, as a function of 
$\kappa$. We verify this prediction by a direct calculation with RG techniques. 

The agreement between these three answers provides a further test of the idea put forward in \cite{mr1,mr2,mr3} that the energy gap in interacting many-fermion 
systems can be understood by using the theory of resurgence and the physics of renormalons.

The paper is organized as follows. In section \ref{sec-review} we review the multi-component Gaudin--Yang model 
and its Bethe ansatz solution. In section \ref{sec-gap-bethe} we calculate the energy gap from the Bethe ansatz equation at weak coupling, extending 
the results of \cite{ko} to the multicomponent case. In section \ref{sec-lo} we study the large order behavior of the perturbative series and we show that it reproduces 
correctly the weak-coupling behavior of the energy gap, in agreement with the conjecture in \cite{mr1}. In section \ref{rg-sec} we compute the beta function of the model by using the RG at two loops, and derive the expression for the gap. Finally, in \ref{sec-conclude} we present some conclusions and prospects for future work. 
There are in addition two Appendices. In the first one we show that ring diagrams lead to an approximate 
expression for the gap which is correct to leading order in the coupling constant, but not to next-to-leading order. In the second Appendix we show that the relativistic 
model obtained in section \ref{rg-sec} by using the approach of \cite{solyom,ls} is closely related to the chiral Gross--Neveu model, and in particular 
leads to the same beta function up to two loops.

\sectiono{The multi-component Gaudin--Yang model and its Bethe ansatz solution}

\label{sec-review}

The Hamiltonian for the Gaudin--Yang model is given by 
\be
H=-\sum_{i=1}^N  \frac{\partial^2}{\partial x_i^2}- 2c \sum_{1\le i<j \le N} \delta(x_i-x_j).  
\ee
We will consider the case of an attractive interaction, which corresponds to a positive coupling constant $c>0$. We also consider the multicomponent 
case, so that each fermion has $\kappa$ possible ``internal" states $|1\rangle, \cdots, |\kappa \rangle$. The number of fermions in the $i$-th internal state $|i \rangle$ 
will be denoted by $N^i$. We will choose the labels of the states, $i=1,\cdots,\kappa$, in such a way that the numbers of particles are ordered as 
$N^1\geq N^2 \geq \cdots \geq N^\kappa$.

The eigenvalue problem for this many-body system can be solved with the Bethe ansatz (BA). We consider the system in an interval of length $L$ and 
we impose periodic boundary conditions. In the case 
$\kappa=2$, the solution was obtained by Gaudin \cite{gaudin} and Yang \cite{yang}. The generalization to arbitrary $\kappa>2$ is due to 
Sutherland \cite{sutherland} and Takahashi \cite{taka}. The solution can be characterized by a system of nested Bethe ansatz equations. In order to 
write down these equations, we introduce 
\be
M_i= \sum_{j=i}^{\kappa-1} N^{j+1} .
\ee
Then, the equations read
\begin{equation}
\ba
\re^{\ri k_i L} &= \prod_{\alpha=1}^{M_1} \frac{k_i-\lambda^{(1)}_\alpha+\ri c'}{k_i-\lambda^{(1)}_\alpha-\ri c'},\quad i=1,\cdots,M_0,\\
\prod_{\eta=1}^{M_l} \frac{\lambda^{(l)}_\alpha-\lambda^{(l)}_\eta+2 \ri c'}{\lambda^{(l)}_\alpha-\lambda^{(l)}_\eta-2 \ri c'} &=-\prod_{\beta=1}^{M_{l-1}}\frac{\lambda_\alpha^{(l)}-\lambda_\beta^{(l-1)}+\ri c'}{\lambda_\alpha^{(l)}-\lambda_\beta^{(l-1)}-\ri c'}\prod_{\delta=1}^{M_{l+1}}\frac{\lambda_\alpha^{(l)}-\lambda_\delta^{(l+1)}+\ri c'}{\lambda_\alpha^{(l)}-\lambda_\delta^{(l+1)}-\ri c'},\\
&\quad \alpha=1,\cdots, M_l, \quad l= 1,\cdots, \kappa-1,
\ea
\label{NBA}
\end{equation}
where $c'=c/2$. The quasi-momenta $k_j$ appearing in (\ref{NBA}) determine the energy eigenvalues through 
\be
E= \sum_{j=1}^N k_j^2, 
\ee
while the Bethe roots $\lambda_\alpha^{(l)}$ are auxiliary variables. 

The solutions $k_j$ to the BA equations form ``strings" in the complex plane, corresponding to bound states of $m$ particles, 
where $1\le m\le \kappa$. The number of bound states with $m$ particles, $N_m$, is related to the numbers of particles in the $i$-th state $N^i$ by 
\be
N_m= N^{m}-N^{m+1}, \quad m=1, \cdots, \kappa-1, 
\ee
and $N_\kappa=N^\kappa$. The ``strings" of quasi-momenta, corresponding to a bound state of $m$ particles labelled by $j=1, \cdots, N_m$, have the form 
\be
k_j^{m,q}=\lambda_j^{m}+\ri(m+1-2q)c'+\CO(\re^{-L}),\quad q=1,\cdots,m.   
\ee
For each set of these $k$'s, one has a set of $m-l$ complex roots $\lambda^{(l)m,q}_j$ at the levels $l = 1,\cdots,m-1$, with the form 
\be
\lambda^{(l)m,q}_j = \lambda_j^{m}+\ri (m-l+1-2q)c'+\CO(\re^{-L}),\quad j = 1,\cdots m-l, \quad l = 1,\cdots, m-1.
\ee
They also share the same real part $\lambda_j^{m}$, which corresponds to the unique real root at level $m-1$. These roots characterize the eigenstate made out 
of $N_m$ bound states of size $1\leq m \leq \kappa$. They 
can be found from the following approximate version of the BA equations (which is correct 
up to exponentially small corrections in $L$) \cite{Lee_2011}:
\begin{equation}
\ba 
&m \lambda_j^m L = 2\pi K_j^m + \sum_{p=1}^{m-1}\sum_{q=p}^\kappa\sum_{l=1}^{N_q} 2\tan^{-1}\left(\frac{\lambda_j^m-\lambda^q_l}{(q+m-2p)c'}\right)+\sum^\kappa_{q=m+1}\sum_{l=1}^{N_q} 2\tan^{-1}\left(\frac{\lambda_j^m-\lambda^q_l}{(q-m)c'}\right),\\
& m = 1,\cdots,\kappa\quad j = 1,\cdots,N_m.
\ea
\label{leeNBA}
\end{equation}
In these equations, 
\be
 K_j^m = - \frac{N_m-1}{2}+j-1.
 \ee
In terms of the roots $\lambda_j^m$, the energy of such a state is given by
\begin{equation}
E(N_1,\cdots,N_\kappa)  =\sum_{m=1}^\kappa \sum_{j=1}^{N_m}m\left( (\lambda_j^m)^2-\frac{\left(m^2-1\right) c ^2}{12}\right).
\label{energynba}
\end{equation}

The ground state of the system is found when all $N=\kappa N_\kappa$ fermions are in $\kappa$ bound-states (in the $\kappa=2$ case, 
these are the Cooper pairs). In that case, (\ref{leeNBA}) reduces to
\begin{equation}
\kappa \lambda^\kappa_j L = 2\pi K_j^\kappa + \sum_{l=1}^{N_\kappa} \sum_{p=1}^{m-1}2\tan^{-1}\left(\frac{\lambda^\kappa_j-\lambda^\kappa_l}{(2\kappa-2p)c'}\right).
\label{NBAgs}
\end{equation}
In the thermodynamic limit 
\be
\label{thl}
L\rightarrow \infty, \qquad N\rightarrow\infty, \qquad {N\over L}=n,
\ee
 the position of the roots becomes a continuous variable $\lambda^\kappa_j\rightarrow \lambda$. The state number $K_j^\kappa\rightarrow K(\lambda)$ gives 
 rise to a state density function $f(\lambda) = L^{-1} \rd K(\lambda) / \rd \lambda$. Taking a derivative of (\ref{NBAgs}) with 
 respect to $\lambda$, we find
\begin{equation}
\frac{\kappa}{2\pi}= f(\lambda)+ \frac{1}{2\pi}\int_{-Q}^Q \rd \lambda' f(\lambda') \sum_{p=1}^{\kappa-1} \frac{2 p c}{c^2 p^2 + (\lambda-\lambda')^2},
\label{pregsc}
\end{equation}
where $Q$ is implicitly defined through
\begin{equation}
\int_{-Q}^Q f(\lambda) \, \rd \lambda  = {n \over \kappa}.
\end{equation}
The ground state energy per unit length is then given by 
\be
\label{e-ground}
E= \kappa \int_{-Q}^Q \left( \lambda^2 - {\kappa^2-1 \over 12} c^2 \right) f(\lambda) \rd \lambda. 
\ee
It is convenient to change variables as 
\be
\theta={ \lambda\over c}, \qquad B={Q \over c}, \qquad \rho(\theta)= \pi f(\lambda). 
\ee
In these variables, the integral equation (\ref{pregsc}) characterizing the ground state reads
\be
\rho(\theta)+\int_{-B}^B \rd \theta' K(\theta-\theta')\rho(\theta') = {\kappa \over 2},
\label{gscont}
\ee
where the kernel can be written in terms of the digamma function as follows: 
\be
K(\theta)=\frac{1}{2\pi} \left( \psi(\ri \theta -\kappa +1)+\psi (-\ri \theta -\kappa +1)-\psi (-\ri \theta )-\psi (\ri \theta ) \right).
\label{conteq}
\ee
The integral equation (\ref{gscont}) was studied in \cite{mr1,mr2} with the techniques developed in \cite{volin, volin-thesis}. Let us introduce the 
dimensionless coupling 
\be
\gamma={c\over  n}.
\ee
Then, from the normalization of the ground state distribution function, 
\be
{1\over  \pi} \int_{-B}^B \rho(\theta) \rd \theta={\kappa \over \gamma},
\ee
one finds the following weak coupling expansion for $B$:
\be
B= \frac{\pi }{\gamma  \kappa }
+{\kappa  \over 2 \pi} \log (\kappa )-{\kappa -1 \over 2 \pi} \left(\log \left(\frac{4 \pi ^2}{\gamma  \kappa }\right)+1\right)
+\CO\left(\gamma \right). 
\label{beta-gamma}
\end{equation}
%


\sectiono{The energy gap from the Bethe ansatz}

\label{sec-gap-bethe}

The Bethe ansatz solution summarized in the previous section makes it possible to calculate the energy gap of the model. 
In the case of the Gaudin--Yang model with $\kappa=2$ components, the gap was first calculated in this way by Krivnov and Ovchinnikov in \cite{ko} 
(see also \cite{zhou-exact}). 
We will now extend this calculation to the case of arbitrary $\kappa$. 

To find the energy gap one has to identify the first excited state, which involves ``breaking'' one of the bound states with $\kappa$ fermions in the ground state. 
At weak coupling, the most favorable process is to produce a free fermion (i.e. a ``1'' bound state) 
and a $\kappa-1$ bound state, out of a $\kappa$ bound state. This can be easily tested numerically, but 
also crudely inspected from (\ref{energynba}) by taking $L\rightarrow\infty$ with $N$ finite, in which case the energy is approximately
\begin{equation}
E(N_1,\cdots,N_\kappa) \approx -\sum_{m=1}^\kappa N_m \left(m\frac{\left(m^2-1\right) c^2}{12}\right).
\end{equation}
Therefore, the energy gap is given by
\begin{equation}
\Delta_\kappa = \Delta_\kappa = E(1,0,\cdots,0,1,N_\kappa-1)- E(0,0,\cdots,0,0,N_\kappa),
\end{equation}
which we will compute in the thermodynamic limit (\ref{thl}). 
We can do this by perturbing the ground state problem.
In the ground state we have $N_\kappa$ bound states characterized by the Bethe roots $\lambda^\kappa_j$, which satisfy the equation (\ref{NBAgs}). 
In the first excited state we have
\begin{equation}
\lambda^1_1 \equiv k,\quad \lambda^{\kappa-1}_1\equiv \Lambda,\quad \bar{\lambda}^\kappa_j \equiv \lambda_j^\kappa+\frac{\xi_j}{L},\quad \bar{K}^{\kappa}_j = K^\kappa_j +\frac{1}{2}. 
\end{equation}
The approximate Bethe ansatz equations (\ref{leeNBA}) give us
\begin{align}
\kappa (L \lambda^\kappa_j+  \xi_j )&= 2\pi K^\kappa_j+\pi + 2\tan^{-1}\left(\frac{\bar{\lambda}^\kappa_j-k}{(\kappa-1)c'}\right)+ \sum_{p=1}^{\kappa-1} 2\tan^{-1}\left(\frac{\bar{\lambda}^\kappa_j-\Lambda}{(2p-1)c'}\right)\\
&+ \sum_{p=1}^{\kappa-1} \sum_{l=1}^{N_\kappa-1}2\tan^{-1}\left(\frac{\bar{\lambda}^\kappa_j-\bar{\lambda}^\kappa_l}{2p c'}\right)\label{NBA0xi},\\
Lk &= 2\tan^{-1}\left(\frac{k-\Lambda}{(\kappa-2)c'}\right)+\sum_{l=1}^{N_\kappa-1} 2\tan^{-1}\left(\frac{k-\bar{\lambda}^\kappa_l}{(\kappa-1)c'}\right),\\
(\kappa-1)L \Lambda &= 2\tan^{-1}\left(\frac{\Lambda-k}{(\kappa-2)c'}\right)+\sum_{l=1}^{N_\kappa-1} \sum_{p=1}^{\kappa-1} 2\tan^{-1}\left(\frac{\Lambda-\bar{\lambda}^\kappa_l}{(2\kappa-2p-1)c'}\right).
\end{align}
The last two equations are easy to solve. In the thermodynamic limit, one has
\begin{align}
k &= 2\int_{-Q}^Q \rd \lambda f(\lambda) \tan^{-1}\left(\frac{k-\lambda}{(\kappa-1)c'}\right)+\CO\left(\frac{1}{L}\right),\\
\Lambda &= \frac{2}{\kappa} \sum_{p=1}^{\kappa-1}\int_{-Q}^Q \rd \lambda f(\lambda) \tan^{-1}\left(\frac{\Lambda-\lambda}{(2\kappa-2p-1)c'}\right)+\CO\left(\frac{1}{L}\right).
\end{align}
Since $\tan^{-1}$ is odd and $f$ is even, $k=\Lambda=0$ solve these equations. 

Let us now consider the first equation in (\ref{NBA0xi}). In the last term in (\ref{NBA0xi}) one needs to expand
\begin{equation}
2\tan^{-1}\left(\frac{\bar{\lambda}^\kappa_j-\bar{\lambda}^\kappa_l}{2 pc'}\right) \sim 2\tan^{-1}\left(\frac{\lambda^\kappa_j-\lambda^\kappa_l}{2 pc'}\right)+\frac{4pc'}{L}\frac{\xi_j-\xi_l}{(2 pc')^2+(\lambda^\kappa_j-\lambda^\kappa_l)^2} + \CO\left(\frac{1}{L^2}\right).
\label{xiexpand}
\end{equation}
Notice that the sum over $l$ in the last term is of order $N \approx L$. Putting (\ref{NBA0xi}), (\ref{xiexpand}) and (\ref{NBAgs}) together we find
\begin{equation}
\ba
\kappa \xi_j = &\pi + 2\tan^{-1}\left(\frac{\lambda^\kappa_j-k}{(\kappa-1)c'}\right)+\sum_{p=1}^{\kappa-1}2\tan^{-1}\left(\frac{\lambda^\kappa_j-\Lambda}{(2p-1)c'}\right)-\sum_{p=1}^{\kappa-1}2\tan^{-1}\left(\frac{\lambda^\kappa_j-\lambda^\kappa_{N_\kappa}}{2pc'}\right)\\&+\sum_{l=1}^{N_\kappa}\sum_{p=1}^{\kappa-1}\frac{4pc'}{L}\frac{\xi_j-\xi_l}{(2pc')^2+(\lambda^\kappa_j-\lambda_l^\kappa)^2},
\ea
\end{equation}
and we can take  the continuum limit
\begin{equation}
\ba
\kappa \xi(\lambda) =& \pi +2\tan^{-1}\left(\frac{\lambda}{(\kappa-1)c'}\right)+ \sum_{p=1}^{\kappa-1}2\tan^{-1}\left(\frac{\lambda}{(2p-1)c'}\right) - \sum_{p=1}^{\kappa-1} 2\tan^{-1}\left(\frac{\lambda-Q}{2pc'}\right)\\
&+ \sum_{p=1}^{\kappa-1}\int_{-Q}^Q \rd\lambda' f(\lambda') \frac{4pc'(\xi(\lambda)-\xi(\lambda'))}{(2pc')^2+(\lambda-\lambda')^2}.
\ea
\end{equation}
The last term in the r.h.s. can be simplified by using (\ref{pregsc}), and one finds
\begin{equation}
\ba
2\pi f(\lambda)\xi(\lambda) =& \pi +2\tan^{-1}\left(\frac{\lambda}{(\kappa-1)c'}\right)+ \sum_{p=1}^{\kappa-1}2\tan^{-1}\left(\frac{\lambda}{(2p-1)c'}\right) - \sum_{p=1}^{\kappa-1} 2\tan^{-1}\left(\frac{\lambda-Q}{2pc'}\right)\\
&- \sum_{p=1}^{\kappa-1}\int_{-Q}^Q \rd\lambda' f(\lambda')\xi(\lambda') \frac{4pc'}{(2pc')^2+(\lambda-\lambda')^2}.
\ea
\end{equation}
%
We change variables to $\theta=\lambda/c$, $B=Q/c$, and introduce the distribution $\Psi(\theta)=f(\lambda)\xi(\lambda)$. We find the following integral 
equation for $\Psi(\theta)$,
\begin{equation}
\ba
2\pi \Psi(\theta) =& \pi +2\tan^{-1}\left(\frac{2\theta}{\kappa-1}\right)+ \sum_{p=1}^{\kappa-1}2\tan^{-1}\left(\frac{2\theta}{2p-1}\right) - \sum_{p=1}^{\kappa-1} 2\tan^{-1}\left(\frac{\theta-B}{p}\right)\\
&- \sum_{p=1}^{\kappa-1}\int_{-B}^B \rd\theta' \Psi(\theta') \frac{2 p}{p^2+(\theta-\theta')^2}.
\ea
\label{eq_Psi}
\end{equation}
This generalizes a similar equation in \cite{ko} for $\kappa=2$, to arbitrary $\kappa$. 

The energy gap is given by
\begin{equation}
\ba
\Delta_\kappa& = k^2 + (\kappa-1)\left(\Lambda^2-\frac{(\kappa-1)^2-1}{12}c^2\right)+\sum_{j=1}^{N_\kappa-1}\kappa\left((\bar{\lambda}^\kappa_j)^2-\frac{\kappa^2-1}{12}c^2\right)\\
&-\sum_{j=1}^{N_\kappa}\kappa\left(\lambda^2_j-\frac{\kappa^2-1}{12}c^2\right),
\ea
\ee
and in the thermodynamic limit we find
\be
\Delta_\kappa= 2 \kappa c^2 \int_{-B}^B \theta \Psi(\theta)\rd \theta-\kappa c^2 B^2 + \frac{\kappa(\kappa-1)}{4}c^2.
\end{equation}

To tackle the integral equation (\ref{eq_Psi}), it is convenient to anti-symmetrize it, as in \cite{ko}. We define the odd function
\begin{equation}
h(\theta)= \frac{\Psi(\theta)-\Psi(-\theta)}{2}-\frac{\text{sgn}(\theta)}{2}.
\end{equation}
If we take into account that 
\begin{equation}
\int_{-B}^B\rd \theta'\frac{p\,\text{sgn}(\theta')}{p^2+(\theta-\theta')^2}=2\tan^{-1}\left(\frac{\theta}{p}\right)-\tan^{-1}\left(\frac{\theta+B}{p}\right)-\tan^{-1}\left(\frac{\theta-B}{p}\right),
\end{equation}
we find that the sum of (\ref{eq_Psi}) with its reflection yields
\begin{equation}
\ba
h(\theta)&+\frac{1}{2\pi} \int_{-B}^B K(\theta-\theta')h(\theta)=\tau_0(\theta)-\frac{1}{2}\text{sgn}(\theta),\\
\tau_0(\theta)&= \frac{1}{\pi}\tan^{-1}\left(\frac{2\theta}{\kappa-1}\right)+\frac{1}{\pi}\sum_{p=1}^{\kappa-1}\tan^{-1}\left(\frac{2\theta}{2p-1}\right) -\frac{1}{\pi}\sum_{p=1}^{\kappa-1}\tan^{-1}\left(\frac{\theta}{p}\right),
\ea
\label{eq_h}
\end{equation}
where the kernel $K(\theta)$ is given in (\ref{conteq}). The energy gap has a simple expression in terms of the function $h(\theta)$, 
\be
\frac{\Delta_\kappa}{c^2}= - 4 \int_{B}^\infty \theta h(\theta)\rd \theta. 
\ee

At large $B$, the integral equation (\ref{eq_h}) can be solved with the techniques first introduced in \cite{griffiths, yang-yang}. 
First, we solve (\ref{eq_h}) in the strict limit $B\rightarrow\infty$. This can be done by taking the Fourier transform of the equation, leading to
\begin{equation}
\tilde{h}_0(\omega)= -\frac{\ri}{\omega}\frac{1-\re^{-|\omega|/2}}{1+\re^{-\kappa|\omega|/2}}\left(1-\re^{-(\kappa-1)|\omega|/2}\right) = -\frac{\ri}{\omega}\sum_{n=-\infty}^\infty\frac{\sin\left(\frac{2\pi}{\kappa}\left(n-\frac{1}{2}\right)\right)}{n-\frac{1}{2}}\frac{1}{\omega-\frac{4\pi \ri}{\kappa}\left(n-\frac{1}{2}\right)}.
\end{equation}
From this representation one can invert the Fourier transform,
\begin{equation}
h_0(\theta)=-\frac{1}{\pi}\tan^{-1}\left(\frac{\sin(\pi/\kappa)}{\sinh(2\pi\theta/\kappa)}\right), 
\end{equation}
and one has, at large $B$, 
\begin{equation}
\label{hap}
h_0(\theta+B) \sim -\frac{2}{\pi} \sin\left(\frac{\pi}{\kappa}\right)\re^{-\frac{2\pi}{\kappa}(\theta+B)}+\CO(\re^{-\frac{4 \pi}{\kappa}B}).
\end{equation}
We want to determine now $h(\theta+B)$. At large $B$, we have $h(\theta+B) \approx r(\theta)$, where $r(\theta)$ satisfies 
the integral equation 
\begin{equation}
\label{req}
r(\theta) = -\frac{2}{\pi} \sin\left(\frac{\pi}{\kappa}\right)\re^{-\frac{2\pi}{\kappa}(\theta+B)}+\int_0^B \rd\theta' R(\theta-\theta')r(\theta). 
\ee
The first term in the r.h.s. of (\ref{req}) is the approximate form of $h_0(\theta+B)$ found in (\ref{hap}), and the kernel is 
given by 
\be
R(\theta) =\frac{1}{2\pi}\int_\IR \rd \omega  \frac{\re^{\ri \omega \theta}\tilde{K}(\omega)}{1+\tilde{K}(\omega)},
\end{equation}
where
\begin{equation}
\tilde{K}(\omega) = \int_\IR \rd \theta \re^{\ri \omega \theta} K(\theta) = \frac{\re^{-\left| \omega \right| }-\re^{-\kappa  \left| \omega \right| }}{1-\re^{-\left| \omega \right| }}.
\end{equation}
One can now use Wiener--Hopf techniques to obtain the Fourier transform of $r(\theta)$
\begin{equation}
\CF_+(\omega)=\int_0^\infty \rd\omega\re^{\ri \omega \theta} r(\theta) = -\frac{2}{\pi}\sin\left(\frac{\pi}{\kappa}\right)\re^{-\frac{2\pi}{\kappa}B}\frac{G_+(\omega)G_+(2\pi\ri/\kappa)}{\frac{2\pi}{\kappa}-\ri\omega},
\end{equation}
where
\begin{equation}
G_+(\omega)= \sqrt{\kappa } \frac{ \Gamma \left(1-\frac{\ri \omega }{2 \pi }\right) }{\Gamma \left(1-\frac{\ri \kappa  \omega }{2 \pi }\right)}\exp \left(\frac{\ri \omega \left(\log \left(-\frac{\ri \omega }{2 \pi }\right)-1\right)}{2 \pi }-\frac{\ri \kappa  \omega  \left(\log \left(-\frac{\ri \kappa  \omega }{2 \pi }\right)-1\right)}{2 \pi }\right).
\end{equation}

In terms of $r(\theta)$, the energy gap at large $B$, which corresponds to weak coupling, is then given by 
\begin{equation}
\label{first-gap}
\ba
\frac{\Delta_\kappa}{c^2}& \approx - 4 B \int_0^\infty r(\theta)\rd\theta=-4B\CF_+(0) \\
&=  \frac{8B}{\pi}\sin\left(\frac{\pi}{\kappa}\right)\frac{\re^{\frac{1}{\kappa }-1} \kappa ^{\frac{1}{\kappa }+1} \Gamma \left(\frac{1}{\kappa }\right)}{2 \pi }\re^{-\frac{2\pi}{\kappa}B}.
\ea
\end{equation}
This constant overall factor can be tested numerically. We have done so for $\kappa=2,3,4,7,8$, finding a relative error of at most $10^{-2}$ using only 10 values of $B$. The result (\ref{first-gap}) generalizes the calculation of \cite{ko} to arbitrary $\kappa$. 

It remains now to express the result (\ref{first-gap}) in terms of $\gamma$. This last step is non-trivial, and in the calculation in \cite{ko} for $\kappa=2$ it involved a constant which had to be determined numerically. An analytic expression for this constant, leading to a complete answer for $\kappa=2$ at next-to-leading 
order in $\gamma$, was obtained in \cite{ls} by an indirect argument, and later confirmed in \cite{wp} (see \cite{frz}). In the case of general $\kappa$, the methods developed in \cite{mr1,mr2} lead to the explicit expression (\ref{beta-gamma}), which make it possible to obtain the analytic form 
of the answer for arbitrary $\kappa$. By using that result, we can finally write
\begin{equation}
\label{gap-final}
\frac{\Delta_\kappa}{E_F}\approx \left( {\kappa \over 2 \pi} \right) ^{2/\kappa }{64 \over \kappa^2 \Gamma \left(1-\frac{1}{\kappa }\right)}
 \gamma ^{1/\kappa } \re^{- \frac{2\pi^2}{\kappa^2}\frac{1}{\gamma}},
\end{equation}
where
\be
E_F= {\pi n^2 \over 4}
\ee
is the Fermi energy of the free one-dimensional Fermi gas. This expression should understood as the leading asymptotic behavior of the gap as $\gamma \rightarrow 0$. 
It can be easily checked that the expression (\ref{gap-final}) agrees, when $\kappa=2$, with the 
results in \cite{ls,wp,frz}. 

The energy gap determines the fundamental non-perturbative scale of the theory. It is exponentially small in $\gamma$, and its prefactor scales 
with $\gamma$ like $\gamma^{1/\kappa}$. We will now see how the main features of this result can be obtained from two different approaches: 
the behavior of perturbation theory at large order, and a RG analysis.

\sectiono{The energy gap from large order behavior}
\label{sec-lo}

The energy gap (\ref{gap-final}) is clearly a non-perturbative effect. It has been known for a long time that 
non-perturbative effects in quantum physics can be often extracted from the large order behavior of the perturbative series (see e.g. \cite{mmbook} for a 
textbook exposition, and \cite{lgzj} for a collection of articles on the subject). Let us suppose that we have a perturbative series of the form, 
\be
\label{formal}
\varphi(z)= \sum_{k \ge 0} a_k z^k. 
\ee
Here, $z$ is the (small) coupling constant of the problem. 
In most examples in quantum theory, the coefficients $a_k$ grow factorially with $k$. More precisely, we have
\be
\label{ak-growth}
a_k \sim {\mu_0 \over 2 \pi} A^{-k-b} \Gamma\left(k+b \right), \qquad k \gg 1, 
\ee
where $A$, $b$ and $\mu_0$ are parameters that characterize the growth of perturbation theory at next-to-leading order in $1/k$. 
This growth leads to an exponentially small, non-perturbative effect of the form 
\be
\label{np-effect}
\mu_0 z^{-b} {\rm e}^{-A/z}, \qquad z\rightarrow 0. 
\ee
Therefore, the parameters in the factorial growth (\ref{ak-growth}) determine the strength of the non-perturbative effect. 
In real examples, these parameters can be extracted numerically from the growth of the perturbative series, and then compared with expectations about the presence of non-perturbative effects. Particularly important are $A$ and $b$, since they determine the leading 
dependence of the non-perturbative effect on the coupling constant $z$. In general, there is a ``minimal" non-perturbative scale in the problem 
\be
\label{min-scale}
\Lambda(z)= z^{-b} \re^{-A/z}
\ee
and a generic non-perturbative effect scales at small $z$ as $\Lambda^d(z)$, where $d$ is often an integer. We note 
that (\ref{min-scale}) is often the leading approximation to the full answer, and it multiplies a power series in $z$.

An illustrative example of the considerations above is the double-well potential in one-dimensional quantum mechanics, of the form 
\be
V(x)= {x^2 \over 2} (1+ x g^{1/2})^2. 
\ee
Here, $g$ can be regarded as a  coupling constant, and the energy levels can be computed as 
formal power series in $g$ by using standard 
stationary perturbation theory. In this potential, the energy gap, i.e. the difference between the ground state energy and the first excited state, is purely non-perturbative in $g$. At leading order, it is given by the scale 
\be
\label{min-dw}
\Lambda(g)= g^{-{1\over 2}} \re^{-{1\over 6g}},  
\ee
and it is due to tunneling between the two classical vacua (in the language of instantons, this is a one-instanton effect). 
One way to extract this scale is to look at the larger order behavior of the perturbative series for the ground-state energy. 
Its coefficients grow as \cite{BPZJ,zjj1}
\be
a_k \sim \Gamma(k+1) 3^{k+1}, \qquad k \gg 1,
\ee
so they lead to a non perturbative scale which is the {\it square} $\Lambda^2(g)$ of the ``minimal" scale (\ref{min-dw}). 

It was conjectured in \cite{mr1} that precisely this phenomenon occurs in Fermi systems with an attractive interaction: the 
large order behavior of the perturbative series for the ground-state energy leads to a non-perturbative scale which is the 
{\it square} of the scale appearing in the energy gap. This was verified for the Gaudin--Yang model with $\kappa=2$ components. We will now 
provide evidence for the same phenomenon in the multi-component case. This in particular will determine a ``minimal" non-perturbative scale 
\be
\label{minimal}
\Lambda(\gamma) = \gamma^{1/\kappa} \re^{-{2 \pi^2 \over \kappa^2 \gamma}}, 
\ee
in agreement with (\ref{gap-final}). 

\begin{figure}
\center
\includegraphics[height=4cm]{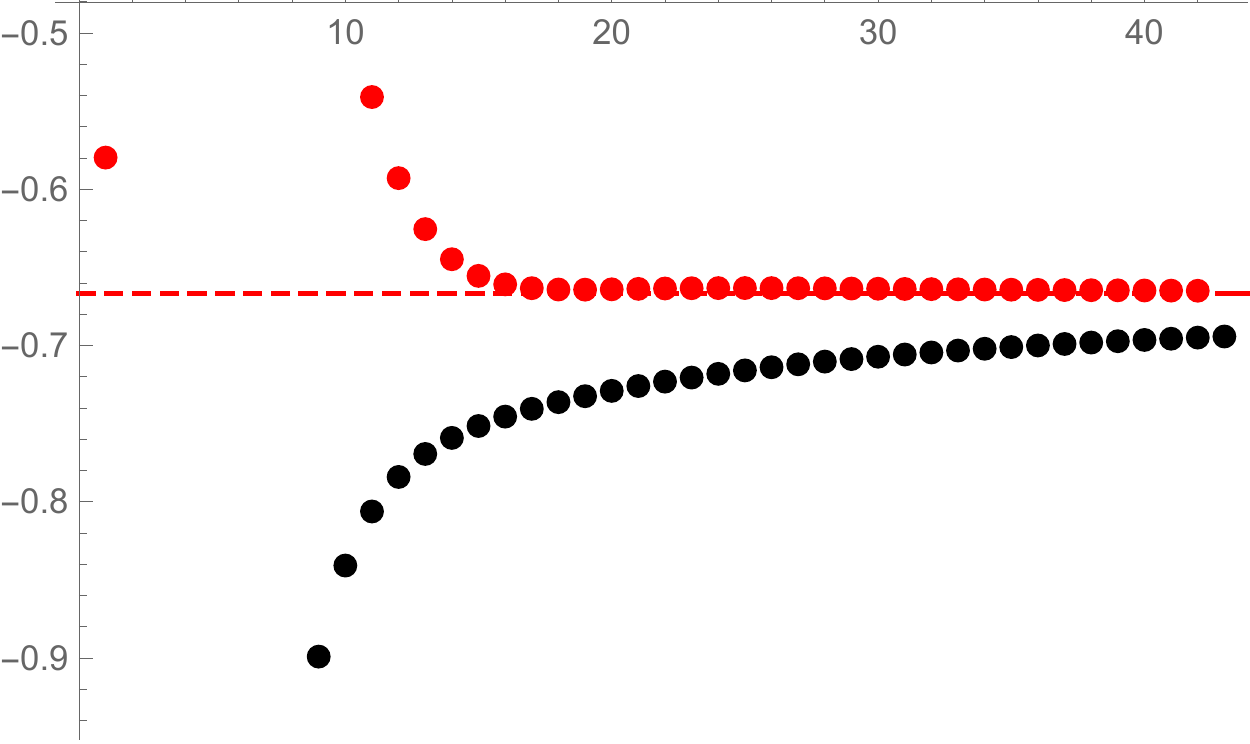} \qquad \qquad \includegraphics[height=4cm]{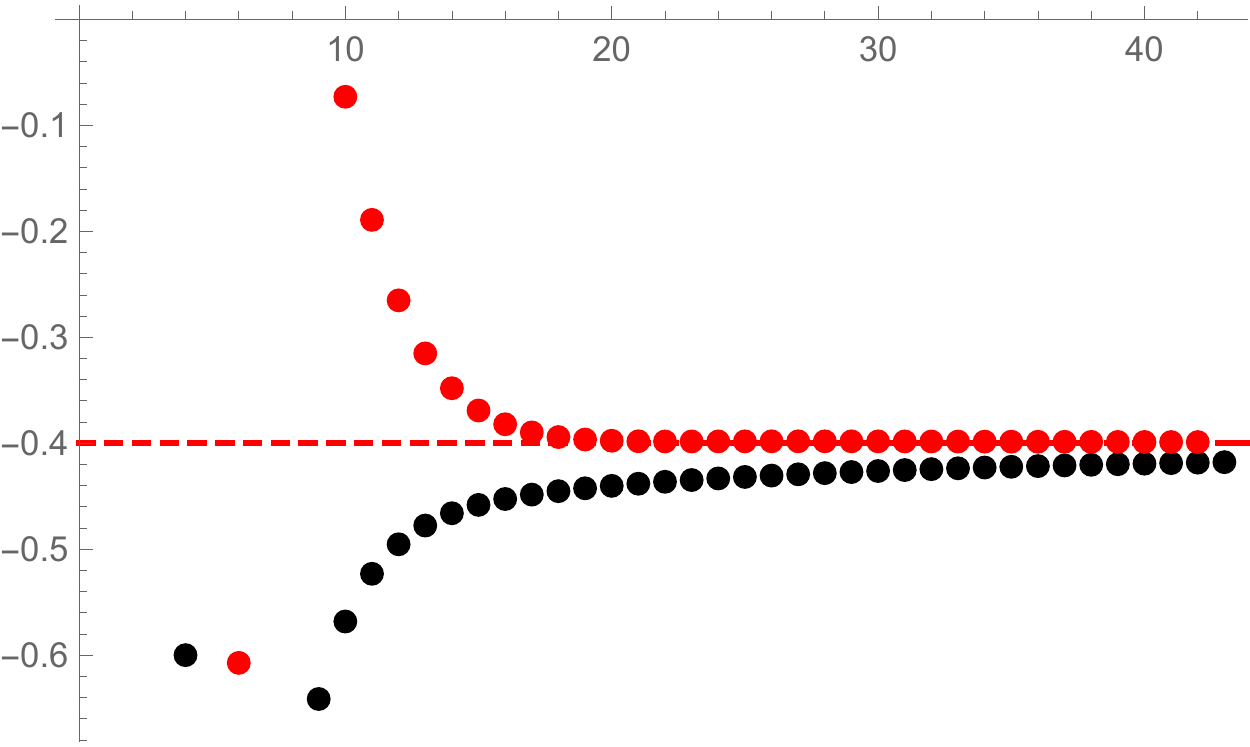}
\caption{The figure in the left (respectively, right) shows the sequence (\ref{tell}) (in black dots) and its 
first Richardson transform (in red dots) for the values $\kappa=3$ (respectively, $\kappa=5$). The horizontal dashed 
line is the expected value $b(\kappa)=-2/\kappa$.}
\label{kappa-bs}
\end{figure}

Following \cite{mr1}, it is useful to introduce the 't Hooft-like coupling 
\be
\label{lam-thooft}
\lambda=  \left( {\kappa \over 2} \right)^2 \gamma 
\ee
and the rescaled ground energy density
\be
\label{elambda}
e(\lambda; \kappa)= {1\over 4} {E/\kappa \over (n/\kappa)^3}, 
\ee
where $E$ is given in (\ref{e-ground}) in terms of the Bethe ansatz solution. This function has the perturbative expansion
\be
\label{e-perturbative}
e(\lambda;\kappa)=\sum_{\ell \ge 0} c_\ell(\kappa) \lambda^\ell. 
\ee
The coefficients $c_\ell(\kappa)$ can be computed systematically by using the algorithm presented in \cite{mr1}. One finds, for the very first orders, 
\be
c_0={\pi^2 \over 12}, \quad c_1=\Delta-1, \quad c_2= {1\over 3}-{\Delta \over 3}, \quad c_3={4 \Delta  (\Delta-1) \zeta(3) \over \pi^4},  
\quad c_4=-{12 \Delta (\Delta-1)^2 \zeta(3) \over \pi^6}, 
\ee
where we have denoted
\be
\Delta={1\over \kappa}. 
\ee
We have computed the first 45 coefficients in (\ref{e-perturbative}), which turn out to be sufficient to study numerically the large order behavior of the 
sequence $c_\ell(\kappa)$. We find,
\be
\label{cell-lob}
c_\ell(\kappa) \sim A^{-\ell -b(\kappa)} \Gamma(\ell+ b(\kappa)), 
\ee
where 
\be
A= \pi^2, \qquad b(\kappa)= -{2\over \kappa}. 
\ee
The numerical procedure to extract these numbers is standard (see e.g. \cite{msw}).  For example, to determine $b(\kappa)$, 
we consider the sequence 
\be
\label{tell}
t_\ell ={A c_{\ell+1} \over c_\ell} - \ell, \qquad \ell \ge 0, 
\ee
which should approach $b(\kappa)$ as $\ell\gg 1$. The convergence of the sequence to the expected value can be accelerated with 
Richardson transforms. Examples of these numerical determinations are shown in \figref{kappa-bs}. 
If we now take into account that the expansion (\ref{e-perturbative}) is done in the 
coupling $\lambda$, and we go back to the coupling $\gamma$, we find that the large order growth leads to the non-perturbative scale
\be
\label{square-gap}
\Lambda^2(\gamma) = \gamma^{2/\kappa} \re^{-{4 \pi^2 \over \kappa^2 \gamma}}, 
\ee
which is precisely the square of (\ref{minimal}). 

As explained in \cite{mr1}, the factorial growth of perturbation theory is due to a 
renormalon effect \cite{beneke}, and ring diagrams explain the exponential dependence in (\ref{square-gap}). 
However, as we check explicitly in Appendix \ref{ring-diags}, the prefactor $\gamma^{2/\kappa}$ is subleading in 
$1/\kappa$ and cannot be explained by ring diagrams only.

\sectiono{The energy gap from the renormalization group}
\label{rg-sec}

As it is well-known, in relativistic asymptotically free theories, the coupling dependence of the non-perturbative scale can be determined, at weak coupling, by a RG analysis. The argument is very simple. Let us assume that we have a running coupling constant $g(\mu)$, 
 depending on a scale $\mu$, and satisfying a RG equation of the form 
\be
\label{rg-eq}
\mu {\rd g \over \rd \mu} =\beta(g)=\beta_0 g^2 + \beta_1 g^3 +\cdots, 
\ee
where $\beta_0<0$. Then, the following quantity 
\be
\label{rginv}
\CI (g)=  \mu \, g(\mu)^{\beta_1/\beta_0^2}{ \rm e}^{{1\over \beta_0 g(\mu)}} 
\exp \left\{ - \int_{g_*}^{g(\mu)} \left( {1\over \beta(\overline g)} -{1\over \beta_0 \overline g^2}+{\beta_1 \over \beta_0^2 \overline g} \right) \rd \overline g \right\}
\ee
is invariant under the RG flow, i.e. it is independent of the scale $\mu$. 
Here $g_\star$ is an arbitrary value, which is equivalent to the freedom of multiplying $\CI$ by an arbitrary $\mu$-independent constant. 
Since $\beta_0<0$, $\CI(g)$ is an exponentially small quantity in the coupling constant, and it can be regarded as the all-orders generalization of the 
``minimal" non-perturbative scale (\ref{min-scale}) for these theories. We note that $A$ in (\ref{min-scale}) is essentially given by the inverse of the 
first coefficient of the beta function, while $b$ involves the first two coefficients $\beta_0$, $\beta_1$. It was pointed out by Parisi \cite{parisi2} that 
the non-perturbative ambiguities due to renormalons are given by integer powers of the scale (\ref{rginv}), and he conjectured that they 
govern the large order behavior of the corresponding perturbative series. 

There are many structural similarities between many-fermion systems with an attractive interaction and asymptotically free 
field theories. One could then use the RG equations to determine the coupling constant dependence of non-perturbative quantities. 
In the case of one-dimensional Fermi systems, this was pointed out by Larkin and Sak in \cite{ls}. In particular, they determined the energy gap in the Gaudin--Yang model with $\kappa=2$ from the RG equations of \cite{m-solyom}. 

In this section we determine the RG equations for the Gaudin--Yang model with arbitrary $\kappa$ and we rederive the non-perturbative scale (\ref{minimal}) . This shows that the connection between non-perturbative 
effects, RG equations and large order behavior in asymptotically free, relativistic field theories, 
also holds in this one-dimensional many-body model. As in \cite{ls}, we will use the RG approach of \cite{m-solyom}, which we will call 
multiplicative renormalization (see \cite{solyom} for a review). 

\begin{figure}[!ht]
\leavevmode
\begin{center}
\includegraphics[width=\columnwidth]{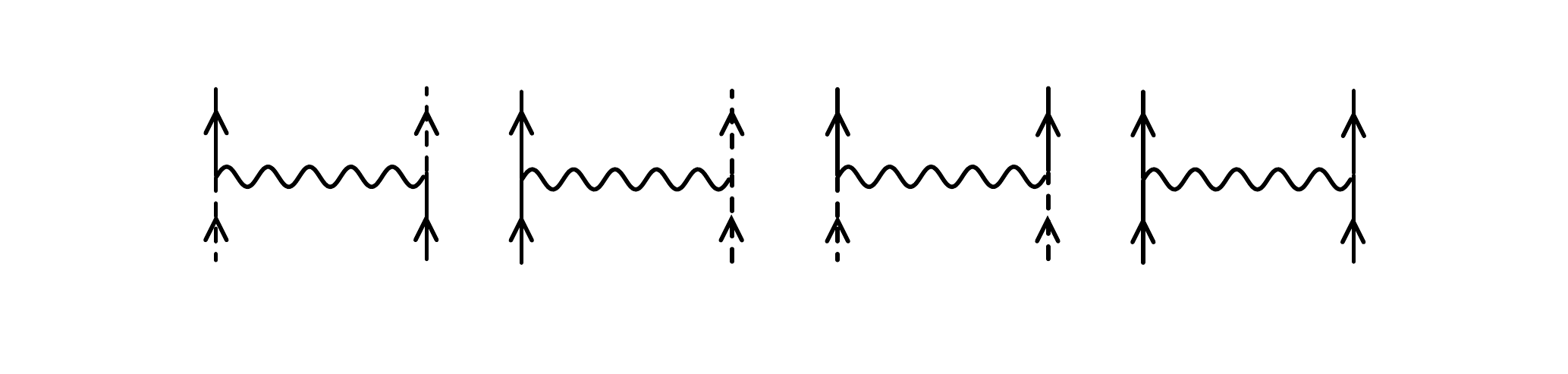}
\end{center}
\caption{The allowed couplings $g_1$, $g_2$, $g_3$ and $g_4$, from left to right. They mix right moving (continuous line) and left moving (dashed line) particles. We only consider $g_1$ and $g_2$, the leftmost couplings. The arrows are taken to be implicit in other diagrams.}
\label{couplings2}
\end{figure}

As is well know, the first step in the multiplicative renormalization procedure in one-dimensions is to linearize the dispersion relation near the Fermi surface. We 
start with a free Hamiltonian 
\begin{equation}
H_0 = \sum_{k,\alpha} \epsilon_k  c^\dagger_{k,\alpha}c_{k,\alpha},
\end{equation}
where $\alpha=1, \cdots, \kappa$. We focus our attention on energies around $k=\pm k_F$, and integrate out modes with 
$\left|k - k_F\right| \gg k_0$ for some cutoff $k_0 \ll k_F$. This leads to a Hamiltonian of the form 
\begin{equation}
H_0 = \sum_{k,\alpha} v_F(k-k_F) a^\dagger_{k,\alpha}a_{k,\alpha}+\sum_{k,\alpha} v_F(-k-k_F) b^\dagger_{k,\alpha}b_{k,\alpha}
\end{equation}
where $a$ and $b$ are annihilation operators for right and left moving particles, respectively, and $v_F$ is the Fermi velocity. The energy 
bandwidth associated to the cutoff $k_0$ is 
given by 
\be
E_0= 2 v_F k_0. 
\ee
We also define the free Green's function for right/left movers as
\begin{equation}
G_\pm(k,\ri \omega) = \frac{1}{\ri \omega \mp k + k_F}
\end{equation}
We can now add interactions which are diagrammatically illustrated in Figure \ref{couplings2},
\be
\label{H-interacting}
\ba
H_I &= \sum_{k_1,k_2,k_3,k_4}\sum_{\alpha,\beta}^\kappa\delta(k_1+k_2-k_3-k_4)\biggl\{ g_1 b_{k_1,\alpha}^\dagger a_{k_2,\beta}^\dagger a_{k_3,\alpha} b_{k_4,\beta} +  g_2 b_{k_1,\alpha}^\dagger a_{k_2,\beta}^\dagger b_{k_3,\alpha} a_{k_4,\beta}\\
&\qquad \qquad + g_3(a_{k_1,\alpha}^\dagger a_{k_2,\beta}^\dagger b_{k_3,\alpha} b_{k_4,\beta}+b_{k_1,\alpha}^\dagger b_{k_2,\beta}^\dagger a_{k_3,\alpha} a_{k_4,\beta})\\
& \qquad \qquad+ g_4 (a_{k_1,\alpha}^\dagger a_{k_2,\beta}^\dagger a_{k_3,\alpha} a_{k_4,\beta}+b_{k_1,\alpha}^\dagger b_{k_2,\beta}^\dagger b_{k_3,\alpha} b_{k_4,\beta})
\biggr\}.
\ea
\ee
Very often the couplings $g_i$ are split into $g_{i \bot}$ and $g_{i \parallel}$ for particles with different/identical spin. 
In the Gaudin--Yang case, the $g_3$ interaction, which corresponds to Umklapp scattering, is not allowed. In the present 
scheme of bandwidth cutoff, the $g_4$ process does not contribute. We will then focus on the couplings $g_{1,2}$. 

The procedure of multiplicative renormalization is based on comparing Green's functions and vertex functions at different values of the cutoffs. 
The working hypothesis is that, once the coupling constants are appropriately adjusted, these functions differ in a multiplicative factor only. 
We have, for the Green's functions,
\begin{equation}
G\left(k,\omega,g'_i,E'_0\right)= z\left(\frac{E'_0}{E_0},g_i\right) G(k,\omega,g_i,E_0).  
\label{Greenscale}
\end{equation}
The vertex or four-point functions are associated to the couplings and related through the equation, 
\begin{equation}
\Gamma'_i(\{k,\omega\},g'_i,E'_0) = z_i^{-1}\left(\frac{E'_0}{E_0},g_i\right) \Gamma_i(\{k,\omega\},g_i,E_0), \qquad i=1,2, 
\label{Vertexscale}
\end{equation}
where $\{k, \omega\}$ denote the four different momenta and frequencies appearing in the vertex. 
This leads to the following renormalization of the coupling constant
\begin{equation}
g'_i =  \frac{z_i\left(\frac{E'_0}{E_0},g_i\right)}{z^2\left(\frac{E'_0}{E_0},g_i\right)} g_i, 
\label{g_ren}
\end{equation}
and to the beta functions 
\be
\label{beta-def}
\beta_i= {\rd g_i (\mu) \over  \rd \log \mu}=\frac{\rd}{\rd \log \mu}\left(\frac{z_i\left(\mu,g_i\right)}{z^2\left(\mu,g_i \right)}\right)\biggr|_{\mu=1} g_i, \qquad i=1,2, 
\ee
where $\mu=E_0'/E_0$. 	

At one loop the procedure is rather simple. Self-energy corrections vanish, so we only need the vertices. 
In order to compute the scaling of $\Gamma_{1,2}$, we must first assign a set of external momenta and frequency. 
Since we are ultimately interested in how the couplings vary with the scale, we can choose one of the external parameters to play the role 
of ``probe scale''. As usual in renormalization, we work under the assumption that we probe energies far below the cut off. 
We choose, following \cite{solyom}, $\omega$, though one could just as well pick $k$ or even the inverse temperature $\beta$. 
A convenient choice of external parameters is proposed in \cite{solyom}. We set the momenta of right/left movers at the Fermi points $\pm k_F$, respectively. The incoming right moving particle has an energy of $3 \ri\omega/2$ while the incoming left moving particle has an energy of $-\ri\omega/2$ and both outgoing particles have the energy $\ri\omega/2$. 

 \begin{figure}[!ht]
\leavevmode
\begin{center}
\includegraphics[width=0.7\columnwidth]{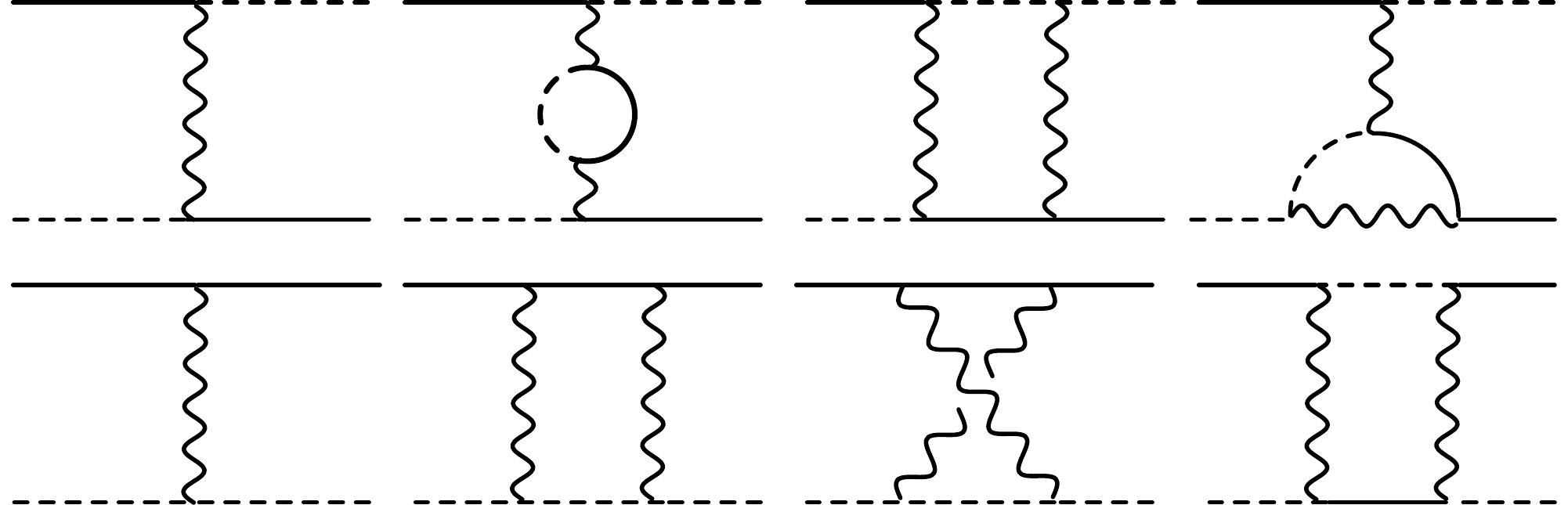}
\end{center}
\caption{One loop correction to $g_1$ (top) and $g_2$ (bottom). The top rightmost diagrams have a multiplicity of $2$ since we can pick the lines to be ingoing/outgoing in two distinct ways.}
\label{vertices1loop}
\end{figure}

At one loop one has the diagrams shown in \figref{vertices1loop}. There are only two types of loop integrals, which correspond to 
the so-called Cooper and Peierls channels, denoted by $J_C(\omega)$ and $J_P(\omega)$, respectively. We can write at first order
\begin{align}
g_1\Gamma_1(\ri\omega,g_i,E_0)&= g_1 - \left(2g_1 g_2 J_C(\omega) + (2g_2g_1-\kappa g_1^2)J_P(\omega)\right)+\cdots,\\
g_2\Gamma_2(\ri\omega,g_i,E_0)&= g_1 - \left((g_1^2+g^2_2) J_C(\omega) + g_2^2 J_P(\omega)\right)+\cdots.
\end{align}
We take $\omega \ll E_0$ to single out the leading logarithmic dependence, and we find 
\begin{equation}
\label{PC_channels}
\ba
J_C(\omega) &= \int_{k_F-k_0}^{k_F+k_0}\frac{\rd q}{2\pi}\int_{-\infty}^{\infty}\frac{\rd \omega'}{2\pi} G_+(q,\ri\omega')G_-(-q,\ri\omega-\ri\omega') 
\approx - \frac{1}{2\pi v_F} \log \left(\frac{\omega }{E_0}\right), \\
J_P(\omega) &= \int_{-k_0}^{+k_0}\frac{\rd q}{2\pi}\int_{-\infty}^{\infty}\frac{\rd \omega'}{2\pi} G_+(q+k_F,\ri\omega')G_-(q-k_F,\ri\omega'-\ri\omega) = - J_C(\omega).\\ 
\ea
\end{equation}
The vertices are 
\begin{equation}
\ba 
g_1\Gamma_1(\ri\omega,g_i,E_0) &= g_1
+\frac{1}{2\pi v_F}\left(\kappa g_1^2\right)\log \left(\frac{\omega }{E_0}\right)
+\cdots\,,\\
g_2\Gamma_2(\ri\omega,g_i,E_0) &= g_2+\frac{1}{2\pi v_F}\left(g_1^2\right)\log \left(\frac{\omega }{E_0}\right)+\cdots.\\
\ea
\end{equation}
From the definition (\ref{Vertexscale}) we read
\begin{equation}
\ba
z_1\left(\frac{E'_0}{E_0},g_i\right)&= \frac{\Gamma_1(\omega,g_i,E_0)}{\Gamma_1(\omega,g'_i,E'_0)}=1+\frac{1}{2\pi v_F}(\kappa g_1) \log\left(\frac{E'_0}{E_0}\right)+\cdots\\
z_2\left(\frac{E'_0}{E_0},g_i\right)&= \frac{\Gamma_2(\omega,g_i,E_0)}{\Gamma_2(\omega,g'_i,E'_0)}=1+\frac{1}{2\pi v_F}\left( \frac{g^2_1}{g_2}\right) \log\left(\frac{E'_0}{E_0}\right)+\cdots
\ea
\end{equation}
where we use $g'_i\approx g_i+\CO(g^2)$. These results are independent of $\omega$, as required by the 
multiplicative renormalization hypothesis. At one loop we find, by using (\ref{g_ren}), 
\begin{align}
g'_1 &= g_1+\frac{g_1^2 \kappa }{2 \pi  v_F} \log \left(\frac{E'_0}{E_0}\right)+\cdots,\qquad 
g'_2 = g_2+\frac{ g_1^2  }{2 \pi  v_F}\log \left(\frac{E'_0}{E_0}\right)+\cdots.
\label{gprime1}
\end{align}

\begin{figure}[!ht]
\leavevmode
\begin{center}
\includegraphics[width=0.65\columnwidth]{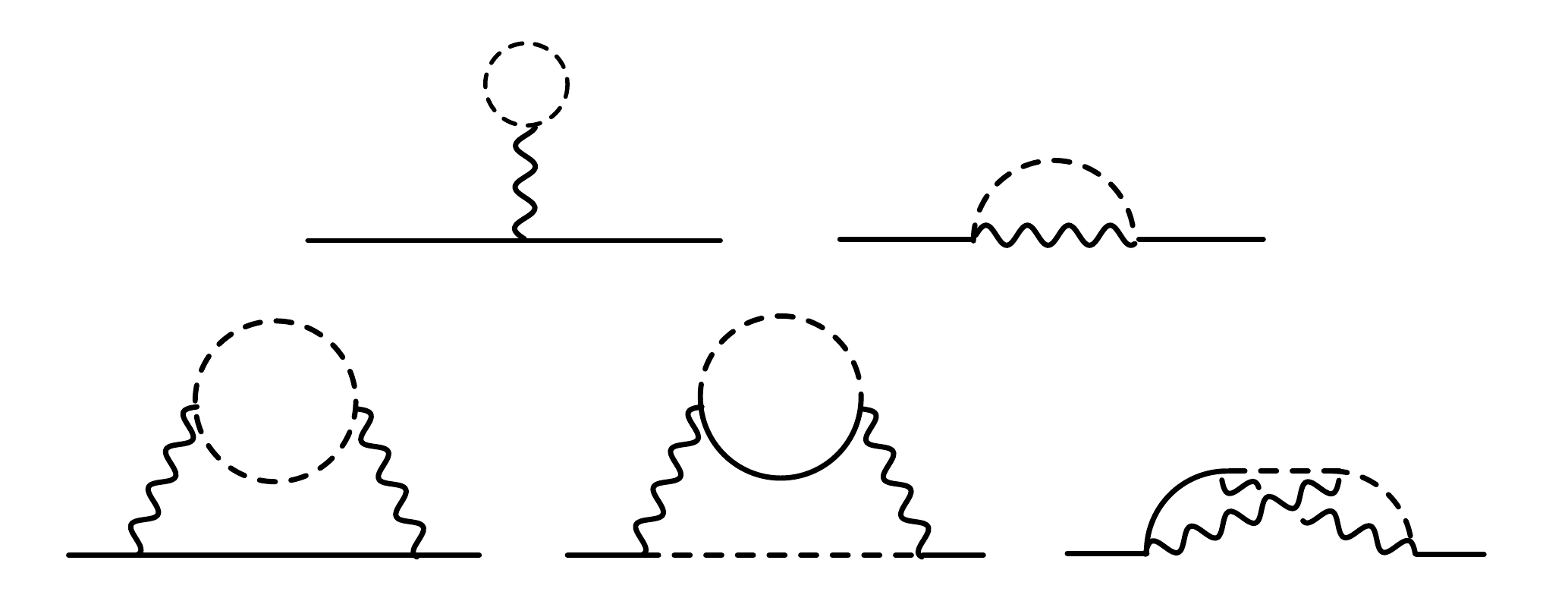}
\end{center}
\caption{Corrections to the right moving self-energy. Note that one-loop contributions are zero. The left moving self-energy is identical.}
\label{propagator}
\end{figure}

The calculation at two loops is more involved. For the self-energy we have the diagrams in \figref{propagator}. 
We take the inflowing momentum and energy to be $k_F+k$ and $\ri\omega$, respectively, and we find 
\begin{equation}
G(k,\ri\omega,g_i,E_0)=\left\{ 1+\left(\kappa g_1^2+\kappa g_2^2 - 2 g_1g_2\right)\frac{\log\left(\frac{\omega}{E_0}\right)}{8\pi^2v_F^2}+\cdots\right\}G_+(k,\ri\omega). 
\label{2loopG}
\end{equation}
Due to (\ref{Greenscale}), $z$ is given by
\begin{equation}
\label{dz}
z\left(\frac{E'_0}{E_0},g_i\right) = \left(\frac{G(k_F,\ri \omega,g_i,E_0)}{G(k_F,\ri \omega,g'_i, E'_0)}\right)^{-1} = 1-\left(\kappa g_1^2+\kappa g_2^2 - 2 g_1g_2\right)\frac{\log\left(\frac{E'_0}{E_0}\right)}{8\pi^2v_F^2}+\cdots.
\end{equation}
%

 \begin{figure}[!ht]
\leavevmode
\begin{center}
\includegraphics[width=\columnwidth]{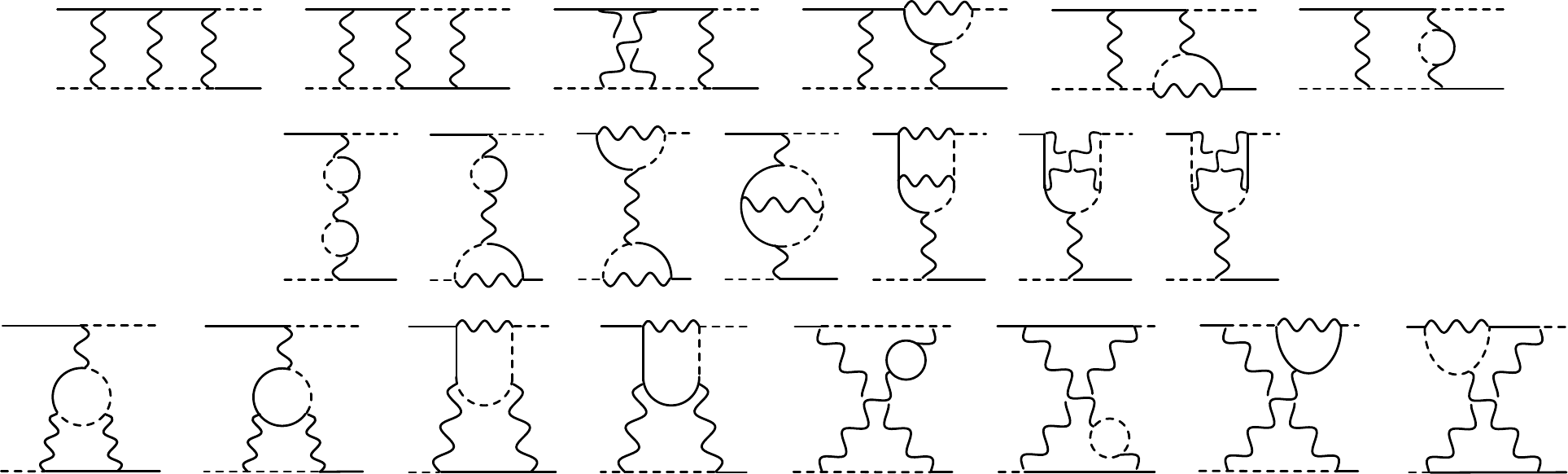}
\end{center}
\caption{Two loop correction to $g_1$. Some diagrams which are distinct over choice of ingoing/outgoing legs have a multiplicity of 2, diagrams with a spin loop have a factor of $-\kappa$.}
\label{twoloopg1}
\end{figure} 

\begin{figure}[!ht]
\leavevmode
\begin{center}
\includegraphics[width=\columnwidth]{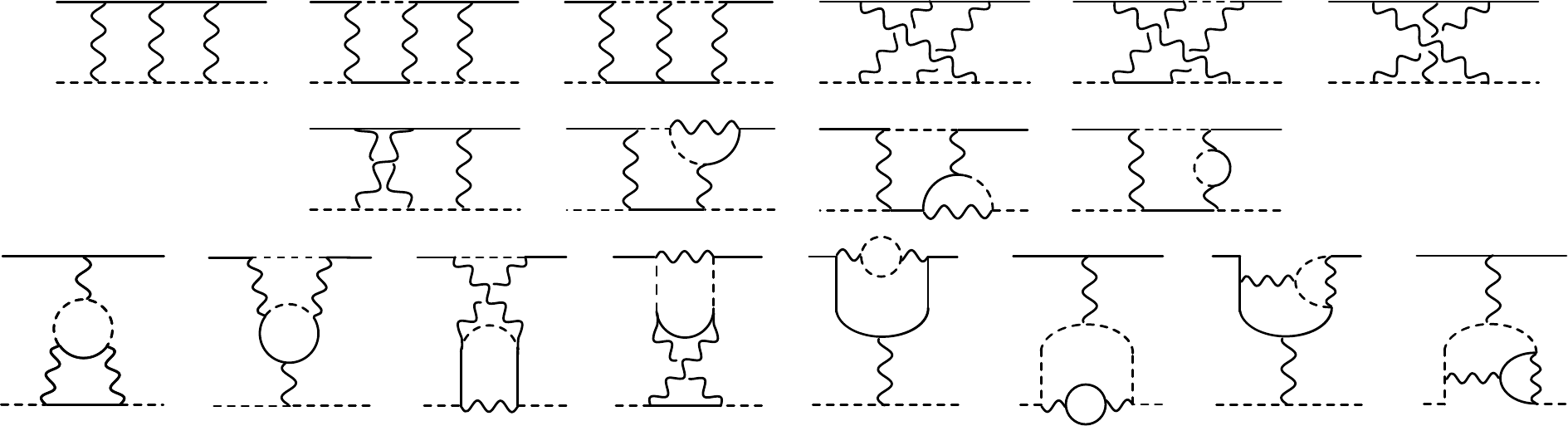}
\end{center}
\caption{Two loop corrections to $g_2$. Some diagrams which are distinct over choice of ingoing/outgoing legs have a 
multiplicity of $2$; diagrams with a spin loop have a factor of $-\kappa$.}
\label{twoloopg2}
\end{figure}

At two loops one finds far more diagrams for the vertices, as detailed in \figref{twoloopg1} and \figref{twoloopg2}. This gives, 
\begin{equation}
\ba 
g_1\Gamma_1(\ri\omega,g_i,E_0) &= g_1
+\frac{1}{2\pi v_F}\left(\kappa g_1^2\right)\log \left(\frac{\omega }{E_0}\right)
+\frac{1}{4\pi^2 v_F^2}\left(2g_1^2g_2-\kappa g_2^2 g_1\right)\log \left(\frac{\omega }{E_0}\right)\\
&+\frac{1}{8\pi^2 v_F^2}\left(2\kappa^2 g_1^3\right)\log^2 \left(\frac{\omega }{E_0}\right)+\cdots \\
g_2\Gamma_2(\ri\omega,g_i,E_0) &=  g_2+\frac{1}{2\pi v_F}\left(g_1^2\right)\log \left(\frac{\omega }{E_0}\right)+\frac{1}{4\pi^2 v_F^2}\left(g_1^3-\kappa g_1^2g_2+2g_2^2g_1-\kappa g_2^3\right)\log \left(\frac{\omega }{E_0}\right)\\&+\frac{1}{8\pi^2 v_F^2}\left(2\kappa g_1^3\right)\log^2 \left(\frac{\omega }{E_0}\right)+\cdots\\
\ea
\label{gamma2loops}
\end{equation}
When $\kappa=2$, the result above agrees with the calculation in \cite{m-solyom,solyom}. From these results we obtain
\begin{equation}
\ba
z_1\left(\frac{E'_0}{E_0},g_i\right)&=  1
+\frac{1}{2\pi v_F}\left(\kappa g_1\right)\log \left(\frac{E'_0}{E_0}\right)
+\frac{1}{4\pi^2 v_F^2}\left(2g_1g_2-\kappa g_2^2 \right)\log \left(\frac{E'_0}{E_0}\right)\\
&+\frac{1}{8\pi^2 v_F^2}\left(2\kappa^2 g_1^2\right)\log^2 \left(\frac{E'_0}{E_0}\right)+\cdots\\
\ea
\label{z1}
\end{equation}
Here we must use the corrections (\ref{gprime1}) to $g'_i$, since they are crucial to cancel $\log(\omega)$ dependencies at order $\CO(g^2)$. 
For $z_2$ we find
\begin{equation}
\ba 
z_2\left(\frac{E'_0}{E_0},g_i\right)
&= 1+\frac{1}{2\pi v_F}\left(\frac{g_1^2}{g_2}\right)\log \left(\frac{E'_0}{E_0}\right)+\frac{1}{4 \pi ^2 v_F^2}\left(\frac{g_1^3}{g_2}- \kappa g_1^2  - \kappa g_2^2  +2 g_2 g_1\right) \log \left(\frac{E'_0}{E_0}\right)\\
&+\frac{1}{8 \pi ^2  v_F^2} \left(2  \kappa \frac{g_1^3}{g_2}\right) \log^2 \left(\frac{E'_0}{E_0}\right)+\cdots\,.
\ea
\label{z2}
\end{equation}
The cancellation of $\log(\omega)$ for $z_1,z_2$ is a non-trivial check of the diagrammatic calculations.

Finally, by assembling the pieces and plugging them in (\ref{g_ren}), we find
\begin{align}
g'_1 &= g_1+\frac{g_1^2 \kappa }{2 \pi  v_F} \log \left(\frac{E'_0}{E_0}\right)+\frac{g_1^3 \kappa   }{4 \pi ^2 v_F^2}\log \left(\frac{E'_0}{E_0}\right)+\frac{g_1^3 \kappa ^2 }{4 \pi ^2 v_F^2}\log^2 \left(\frac{E'_0}{E_0}\right)+\cdots,\\
g'_2 &= g_2+\frac{ g_1^2  }{2 \pi  v_F}\log \left(\frac{E'_0}{E_0}\right)+\frac{g_1^3  }{4 \pi ^2 v_F^2}\log \left(\frac{E'_0}{E_0}\right)+ \frac{  g_1^3 \kappa}{4 \pi ^2 v_F^2}\log^2 \left(\frac{E'_0}{E_0}\right)+\cdots\,.
\label{gprime2}
\end{align}
We are now ready to calculate the beta functions for the couplings $g_1$, $g_2$. By using (\ref{z1}), (\ref{z2}) and (\ref{dz}) in (\ref{beta-def}), we obtain
\begin{align}
\beta_1 &= \frac{\kappa}{2\pi v_F} g_1^2 + \frac{ \kappa   }{4 \pi ^2 v_F^2}g_1^3 +\cdots\,,\\
\beta_2 &= \frac{1}{2\pi v_F} g_1^2 + \frac{ 1   }{4 \pi ^2 v_F^2}g_1^3 +\cdots\,.
\end{align}
These beta functions agree with a similar calculation in the $SU(\kappa)$ Hubbard model in \cite{solyom-kappa}. 

With all these results, we can now calculate the gap in the attractive regime $g_1=-2c < 0$. We introduce 
\be
\label{coupling}
\bar g_1 =-\frac{g_1}{\pi v_F}= {\kappa \gamma \over \pi^2}, 
\ee
and we find that the beta function for $\bar g_1$ is of the form (\ref{rg-eq}) with:
\be
\beta_0= -\frac{\kappa}{2},\qquad \beta_1 = \frac{\kappa}{4}. 
\end{equation}
By using these values and (\ref{coupling}), we find that the RG invariant scale (\ref{rginv}) agrees precisely with (\ref{minimal}). 

As noted in \cite{mr3}, the beta function $\beta_1$ coincides with the one of the 
chiral Gross--Neveu model \cite{gross-neveu}\footnote{We thank Philippe Lecheminant for 
pointing out the relation to the chiral Gross--Neveu model.}. In fact, it can be shown explicitly 
that the Hamiltonian $H_0+H_I$, where only the couplings $g_1$, $g_2$ are taken into account, is a particular case of the chiral Gross--Neveu model. 
This explains the relationship between the beta functions. We give some details of this equivalence in Appendix \ref{cgn-rel}.

\sectiono{Conclusions}
\label{sec-conclude}

In this paper we have extended and deepened the connection found in \cite{mr1, mr2, mr3} between the energy gap, the large order 
behavior of perturbation theory, and renormalons, in one-dimensional models of many-body fermions with an attractive interaction. 
We have seen that the weak-coupling behavior of the energy gap in the multi-component Gaudin--Yang model 
can be predicted from the large order behavior of the 
perturbative series for the ground state energy. This series diverges factorially due to renormalon diagrams. When the number of components is large, the leading renormalon diagrams (which turn our to be ring diagrams) 
reproduce correctly the exponential term in the energy gap. Moreover, as in asymptotically free theories in two dimensions, the leading and sub-leading terms in the large order behavior can be obtained from the beta function of the theory, as computed in the relativistic approximation near the Fermi points. 
This also implies a connection between the gap and the beta function, noted long ago in \cite{ls}. 
In order to establish these relationships, we have performed a detailed calculation of the energy gap directly 
from the Bethe ansatz solution in the multicomponent case, generalizing in this way the results of \cite{ko} for $\kappa=2$. 

Although the integrability of the model makes it possible to test our ideas in detail, the connection we have found should be valid 
more generally. For example, the results of this paper, combined with the ones in \cite{mr3}, suggest that the energy gap in the 
multi-component Hubbard model (which is not integrable) is given by 
\be
\label{delta-conj}
\Delta \approx u^{1/\kappa} \exp \left(- { 2 \pi \over \kappa u} \sin \left( {\pi n \over \kappa} \right) \right), \qquad u \rightarrow 0, 
\ee
where $u$ is the coupling constant and $n$ is the density (see \cite{mr3} for more details and clarifications on the notation). The exponent appearing in this expression can be interpreted as due to the contribution of renormalon diagrams dominating in the large $\kappa$ limit. In QCD, 
renormalons have been 
instrumental in determining non-perturbative scales \cite{beneke}, and it is gratifying that the same principles shed 
light on the energy gap of many-fermion systems. 

There are various avenues open by this investigation. One important issue would be to understand systematically the 
corrections to the results presented in this paper. As we have mentioned, the ``minimal" scale (\ref{min-scale}) is the 
leading approximation to a fully-fledge trans-series, and it multiplies a power series in the coupling constant. 
As emphasized in this paper, these subleading corrections can in principle be computed by following any of the three roads 
we have considered. We could for example use the Bethe ansatz equations, we could determine them from the subleading contributions to the large order behavior, and we could try to understand them from the beta function, by including higher loops and higher modes. It would be also 
interesting to connect these corrections to the behavior of diagrams. 
In fact, this should be done already to reproduce the prefactor $\gamma^{2/\kappa}$ in (\ref{square-gap}). It might be possible to do this 
by considering diagrams which are subleading in the large $\kappa$ expansion. 

Another interesting avenue is to find a description of the model in the $1/\kappa$ expansion, along the lines of what was done for the principal 
chiral field in \cite{fkw1, fkw,ksz}. This might require to study a regime of the model in which different bound states 
are present in a prescribed way, as in \cite{fkw1,fkw}. We have found encouraging indications that the Bethe ansatz equations for 
the multi-component Gauding--Yang model might simplify in an appropriate 
large $\kappa$ regime, but more work is needed.   

As mentioned in our previous papers \cite{mr1, mr2, mr3}, a fundamental issue is to find a first-principles procedure to calculate the 
energy gap from the path integral, by some generalization of perturbation theory that takes into account renormalon physics. 
In QCD, such a procedure is provided, for some observables, by the OPE, combined with the existence of non-trivial vacuum condensates. 
It would be fascinating to extend these methods to non-relativistic models like the one studied in this paper.

\section*{Acknowledgements}
We would like to thank Thierry Giamarchi, Wilhelm Zwerger and Philippe Lecheminant for useful 
discussions and correspondence. This work has been supported in part by the Fonds National Suisse, 
subsidy 200020-175539, by the NCCR 51NF40-182902 ``The Mathematics of Physics'' (SwissMAP), 
and by the ERC-SyG project ``Recursive and Exact New Quantum Theory" (ReNewQuantum), which received funding from the European Research Council (ERC) under the European Union's Horizon 2020 research and innovation program, 
grant agreement No. 810573.

\appendix 

\sectiono{Non-perturbative scale from ring diagrams}
\label{ring-diags}
As we have already argued in \cite{mr1}, the large order behavior of the perturbative series (\ref{cell-lob}) is due to renormalon 
diagrams. In order to study these diagrams in a systematic way, it is useful to do a large $\kappa$ expansion which isolates 
the most important type of diagrams (in QED or QCD, this is done by taking a large number of flavours \cite{beneke}). 
In the case of the multi-component Gaudin--Yang model, the large $\kappa$ limit selects the so-called ring diagrams (see 
e.g. \cite{hf,coleman}). It was shown in \cite{mr1} that ring diagrams lead to the correct exponential term in (\ref{square-gap}). 
However, as we will briefly show here, they do not lead to the correct prefactor, which is subleading in the $1/\kappa$ expansion.

The ground state energy for the Gaudin-Yang model with $\kappa$ spin components (\ref{elambda}) has a $1/\kappa$ expansion of the form 
\be
e(\lambda;\kappa)= e_0(\lambda) +{1\over \kappa} e_1(\lambda)+ \cdots
\ee
In this equation, $\lambda$ is the 't Hooft parameter (\ref{lam-thooft}), $e_0(\lambda)$ is the free gas result plus the Hartree term, while $e_1(\lambda)$ is given by a resummation of ring diagrams \cite{mr1}:
\be
\label{e1-int}
e_1(\lambda)= \lambda - {\pi \over 4} \int_0^\infty \rd y \, y \int_0^\infty \rd \nu \, \left[ \log \left( 1-{\kappa^2 \gamma  \over \pi^2} F( y, \nu)\right)+ {\kappa^2 \gamma  \over \pi^2} F( y,\nu) \right], 
\ee
where
\be
F( y,\nu)={1\over 2 y} \log \left( {(y/2+1)^2 + \nu^2 \over (y/2-1)^2 + \nu^2 }  \right). 
\ee
As noted in \cite{mr-new, mr3} in similar situations, the integral (\ref{e1-int}) has an exponentially small imaginary piece 
which must be cancelled by non-perturbative effects not captured by the diagrammatic expansion. We can then reconstruct these non-perturbative effects by properly expanding the imaginary part of the integral, which occurs when the argument of the $\log$ becomes negative. The condition 
\be
1-{\kappa^2 \gamma  \over \pi^2} F(y,\nu, \gamma)<0
\ee
defines a compact region $\CR$ in the first quadrant of the $(y, \nu)$ plane. The region is delimited by the curve defined by the equation
\be
\nu^2 = {-(y-2)^2+ \re^{-2 \pi^2 y/\kappa^2 \gamma} (y+2)^2 \over 4 \left(1-\re^{-2\pi^2 y/\kappa^2 \gamma} \right)}. 
\label{eqnu}
\ee
Since $\nu\geq 0$, we must find the limits of integration $y=y_\pm$ where the boundary line crosses the real axis. Let us define the non-perturbative parameter 
\be
\alpha=\re^{-2\pi^2 /\kappa^2 \gamma}
\ee
and let us change variables from $y$ to $u$, where 
\be
y = 2 + 4 \alpha u.
\ee
 The equation for the endpoints $u_\pm$ is
\begin{equation}
\re^{-2\alpha\log\alpha u_\pm}\mp \left(\frac{1}{u_\pm}+\alpha\right)=0, 
\label{eq_upm}
\end{equation}
which can be easily solved in a power series expansion in the two variables $\alpha$, $\log \, \alpha$ (this is a simple example of a 
trans-series, see \cite{ss,mmlargen,abs}). For the first few orders we find
\begin{equation}
\ba
u_\pm &=\pm 1+\alpha  (2 \log (\alpha )+1)\pm \alpha ^2 \left(6 \log ^2(\alpha )+6 \log (\alpha )+1\right)+\cdots\\
\ea
\end{equation}
To determine the imaginary part of $e_1(\lambda)$, we have to calculate 
\begin{equation}
\int_{u_-}^{u_+} 4\alpha(2+4\alpha u)\nu(u)\rd u .
\label{nuint}
\end{equation}
This can be done by expanding the integrand into factors of $(u-u_+)^m(u-u_-)^k$ and $u(u-u_+)^m(u-u_-)^k$ 
at each order in $\alpha$ before performing the integration, 
and then resuming at each order in $\alpha$ the resulting polynomials in $u_+$ and $u_-$. When all this is done, we obtain the 
following expansion for the imaginary part of $e_1(\lambda)$:
\begin{equation}
\ba 
{\rm Im}\, e_1(\lambda)&= 2 \pi ^2 \re^{- \pi^2 /\lambda }
+\frac{8 \pi^2\re^{-2 \pi^2 /\lambda} }{ \lambda^2}\left(\lambda ^2- \frac{3\pi^2}{2}\lambda +\frac{\pi^4}{2}\right)\\
&
+\frac{6 \pi ^{2} \re^{-3 \pi^2 /\lambda}}{ \lambda ^4} \left(3 \lambda ^4-14\pi^2\lambda^3+ 21\pi^4 \lambda^2- 12\pi^6 \lambda + 
\frac{9\pi^8}{4}\right) +\cdots
\ea
\label{trans}
\end{equation}
The leading, exponentially small effect has the correct exponent to match (\ref{square-gap}), but not the correct prefactor. A similar 
phenomenon was found in the Hubbard model in \cite{mr3}. This is due to the fact that ring diagrams capture the diagrammatric structure at the first non-trivial order
in the $1/\kappa$ expansion, while the prefactor $\gamma^{1/\kappa}$ is subleading in $1/\kappa$. By considering renormalon diagrams of order $1/\kappa^2$ one might be able to reproduce this prefactor\footnote{Non-trivial prefactors appearing in energy gaps can sometimes be reproduced by renormalon calculations. An 
example of this occurs in the the two-dimensional model analyzed in \cite{mr-new}. In this model, the mass gap at large $N$ obtained in \cite{mr-new} 
by a renormalon calculation around the perturbative vacuum matches precisely a large $N$ calculation at the 
non-perturbative vacuum \cite{serone-gap}. This includes not only the correct prefactor, but also an infinite series of 
exponentially small corrections given by a Lambert function.}.

An interesting application of the above calculation is a precise formula for the large order behavior of the coefficients 
$c_\ell^{(1)}$ in the perturbative expansion of $e_1(\lambda)$:
\be
e_1(\lambda)= \sum_{ \ell \ge 0} c_\ell^{(1)} \lambda^\ell. 
\ee
These coefficients appear in the $1/\kappa$ expansion of the coefficients $c_\ell(\kappa)$ of (\ref{e-perturbative}):
\be
c_\ell(\kappa)=  c_\ell^{(0)}+{1\over \kappa} c_\ell^{(1)}+ \cdots
\ee
If we write 
\be
{\rm Im}\, e_1(\lambda)= \sum_{j\ge 1} \sum_{i=0}^{2j-2} a_{j,i} \lambda^{-i} \re^{-j \pi^2/\lambda} 
\ee
we find 
\be
\label{lo-csub}
 c_\ell^{(1)} \sim -\sum_{j\ge 1} \sum_{i=0}^{2j-2} {\Gamma(\ell+i) \over (\pi^2 j)^{\ell + i}} a_{j,i}. 
 \ee
By appropriately truncating the sum over $j$, we can obtain from (\ref{lo-csub}) very accurate values for the perturbative coefficients $c_\ell^{(1)}$.

\sectiono{Relation to the chiral Gross--Neveu model}
\label{cgn-rel}

The relativistic model we have used in our RG analysis turns out to be closely related to the Thirring model and, more precisely, to the chiral Gross--Neveu model 
(similar relations have been pointed out in \cite{woy-1, woy-2}). To see this, we consider the general form of the Thirring Lagrangian, given by
\begin{equation}
\CL = \ri \bar{\Psi}\slashed{\partial}\Psi- \frac{1}{2}g J_\mu^\alpha J^{\mu\alpha},\quad J_\mu^\alpha = \sum_j \bar{\Psi}_j \gamma_\mu T^\alpha \Psi_j,
\end{equation}
where $\Psi_{\alpha,j}$ ($j= 1,\cdots, N_f$ and $\alpha=1,\cdots, N_c$) is a $N_f$ dimensional vector of Dirac spinors in an $N_c$ dimensional representation of a compact Lie group $G$, and $T^\alpha,\, \alpha= 1,\cdots,  \text{dim}(G)$ are a basis for the representations of its Lie algebra such that $\text{Tr}[T^\alpha T^\beta] = \frac{1}{2} \delta^{\alpha \beta}$. In this section we reserve the term ``spin'' for spinor indices and use ``colour'' for the $SU(N)$ symmetry.

We are interested in the case $N_f=1$ and $G= SU(N)$, with the Dirac spinor in the fundamental representation ($N_c=N$). This is the matter content 
of the chiral Gross--Neveu model (see e.g. \cite{forgacs-chiral}). We work in 1+1 dimensions with signature $(-,+)$. Explicitly, we use 
the following Dirac spinor conventions, where the chiral matrix is  labelled $\gamma_3$,
\begin{equation}
\Psi = \begin{pmatrix}
\psi_+ \\ \psi_-
\end{pmatrix},\quad
\bar{\Psi} = \ri \Psi^\dagger \gamma_0,\quad 
\gamma_0 = \begin{pmatrix}
0&1\\ 1&0
\end{pmatrix},\quad
\gamma_1 = \begin{pmatrix}
0&-1\\ 1&0
\end{pmatrix},\quad
\gamma_3 = \begin{pmatrix}
1&0\\ 0&-1
\end{pmatrix},\quad
\gamma_3 = \gamma_0\gamma_1,
\end{equation}
and the colour-space index is suppressed for the spinors. The kinetic term is easily expanded in the above convention as
\begin{equation}
\ri \bar{\Psi}\slashed{\partial}\Psi = - \psi^*_+(\partial_0-\partial_1)\psi_+ - \psi^*_-(\partial_0+\partial_1)\psi_-,
\end{equation}
and they match the left/right moving modes from the two Fermi points considred in section \ref{rg-sec} (up to a rescaling of the spatial direction by $v_F$).

In order to compare with the interactions in (\ref{H-interacting}), we need to expand the vertex. It is useful to use two different Fierz identities. The first one is the Fierz identity in the Clifford algebra of two spacetime dimensions,
\begin{equation}
(\gamma_\mu)_\alpha^\beta (\gamma^\mu)_\gamma^\delta  = (\mathbb{I})_\alpha^\delta(\mathbb{I})_\gamma^\beta- (\gamma_3)_\alpha^\delta(\gamma_3)_\gamma^\beta,
\label{fierzspin}
\end{equation}
and the second one is the Fierz identity in the $SU(N)$ Lie algebra
\begin{equation}
(T^\alpha)_{ab}(T^\alpha)_{cd} = \frac{1}{2}\left(\delta_{ad}\delta_{cb}-\frac{1}{N}\delta_{ab}\delta_{cd}\right).
\label{fierzcolour}
\end{equation}
Using these two together we get, after some simple algebra,
\begin{equation}
\label{int-JJ}
\ba
-\frac{1}{2} g J_\mu^\alpha J^{\mu\alpha} &= \frac{g}{4}\left((\bar{\Psi}\cdot\Psi)^2-(\bar{\Psi}\cdot\gamma_3\Psi)^2 +\frac{1}{N}(\bar{\Psi}\cdot\gamma^\mu\Psi)(\bar{\Psi}\cdot\gamma_\mu\Psi)\right)
\\&= -g(\psi^*_+ \cdot \psi_-)( \psi^*_- \cdot \psi_+) - \frac{g}{N}(\psi^*_+ \cdot \psi_+)(\psi^*_- \cdot \psi_-),
\ea
\end{equation}
where the inner products show explicitly the colour index sums. We can already identify the two vertices $g_{1,2}$ in section \ref{rg-sec}
as $g_1 \propto -g$ and $g_2 \propto -g/N$. Note that, according to our results in section \ref{rg-sec}, $g_1-Ng_2$ is RG invariant. Here we find 
an additional perspective on this fact: this combination is forced to be zero due to Lorentz and 
$SU(N)$ invariance of the relativistic Lagrangian. (\ref{int-JJ}) is also the interaction term for the chiral 
Gross--Neveu model in \cite{forgacs-chiral} with $g'=0$. By using the results in \cite{destri} it is also possible to show in detail that the 
calculation of the beta function in section \ref{rg-sec} is identical to the one for the coupling $g$ in the chiral Gross--Neveu model.

\bibliographystyle{JHEP}

\linespread{0.6}
\bibliography{multi-gy}

\end{document}